# A diffraction approach to assess the elastic properties of a nylon string

Francisco M. Muñoz-Pérez, **Adrián Garmendía-Martínez**, Juan C. Castro-Palacio, Vicente Ferrando, Juan A. Monsoriu

Centro de Tecnologías Físicas, Universitat Politècnica de València, 46022 València, Spain.

Corresponding author: jmonsori@fis.upv.es

Abstract

The Young's modulus of a nylon string has been determined experimentally by combining elasticity theory and wave optics. A diffraction experiment has been setup to determine the change in the string diameter for different tensile forces applied by means of hanging weights. A linear elasticity model has been used to calculate the Young's modulus. A simple method has been provided which could be used as the basis of an experiment for introductory university level.

## 1. Introduction

There are many approaches to determine the Young's modulus with interest for Physics teaching [1-11]. In one of the early works published in in 1965, [1] the authors used a brass wire which was anchored to hanging weights. The Cenco-Cowan extensometer, designed for measuring the thermal expansion of a tube, was employed to measure the extension of the wire with considerably accuracy.

In 1973 [2], the Young`s modulus of a wire was determined dynamically using its mechanical resonant frequencies. Dynamic bending vibration measurements were also carried out in Ref. [3] to calculate the Young's modulus for samples of tungsten, copper, and aluminium. Flores-Maldonado et al. in 1987 [5] followed a similar idea to measure the Young's modulus of steel and balsa wood.

The Young's modulus can also be determined from the speed of sound in metals. Authors in Ref. [6] used the method of singing rods [7] to determine the speed of sound, which combined with the density, allows for the calculation of the Young's modulus. This method has been revisited in 2015 [9] by calculating the speed of sound in a rod using a standing wave recorded by a microphone.

In Ref. [10] the length dependence of the audio frequency of a guitar string is used to determine its Young`s modulus. In a very recent article [11], authors used an Arduino sensor to measure the bending frequency of a metal ruler and thus determine the Young's modulus for a harmonic motion of the ruler.

In this work, elasticity theory is combined with wave optics to determine the Young's modulus of a nylon string. The changes in the diameter of the nylon string for different tensile forces applied by means of hanging weights was determined from direct measurements on the diffraction spectrum of the string, when it is illuminated with a standard power He-Ne laser. A linear elasticity model is used to obtain the Young's modulus of the string.

## 2. Physics model

According to Hooke's law the deformations can be described in first-oder approximation as linear to the applied forces [12]. This is the case for ellasticity,

$$\sigma = E \frac{\Delta L}{L_0}, \tag{1}$$

where the Young's modulus is the coefficient of proportionality between the tensile force per area unit and the unit deformation of the material along the direction of the force ($\Delta L/L_0$), $L_0$ being the length at rest.

Similarly, the Poisson's ratio is a measure of the linear relationship between the transverse and the axial strains ($\varepsilon_{trans}$ and $\varepsilon_{axial}$ respectively), that is, the ratio between the deformations of the material perpendicular ($\Delta d/d_0$, $d_0$ being the transverse width at rest) and parallel to the direction along which the force is applied,

$$\nu = \frac{d\varepsilon_{trans}}{d\varepsilon_{axial}} = -\frac{\Delta d}{d_0}\bigg/\frac{\Delta L}{L_0}. \tag{2}$$

Combining equations (1) and (2),

$$\frac{\Delta d}{d_0} = -\nu \frac{\sigma}{E}. \tag{3}$$

Specifically, for the experiment described in this work, the normal tension can be expressed as:

$$\sigma = \frac{mg}{A} = \frac{mg}{\pi d_0^2/4}, \tag{4}$$

where $A = \pi d_0^2/4$ is the area of the cross section of the nylon string, and $m$ the hanging weight used to apply the tensile force.

Substituting Eq. (4) in Eq. (3) and expressing $\Delta d = d - d_0$, one obtains a linear relationship between the resulting diameter of the nylon string and the hanging weight:

$$d = d_0 - \frac{4\nu g}{\pi d_0 E}m \tag{5}$$

This is the main working expression for the results in this work.

For thin enough nylon strings, its diameter can be determined by a diffraction experiment. Indeed, this work combines the topics of elasticity and wave optics to determine the Young's modulus. The resulting diameter of the string for each of the hanging weights is determined from a diffraction experiment. According to Babinet's principle, complementary objects will form identical diffraction patterns. In this case, shining a laser on the string will produce the same diffraction pattern as a single slit of the same width [13,14]. Destructive interferences then occur at distances from the centre of the diffraction pattern such that the positions of the minima are given by the following expression [12],

$$d\frac{\Delta y}{D} = n\lambda, \tag{6}$$

where $D$ is the distance between the string and the observation screen, $\lambda$ the laser wavelength, and $n$ an integer.

## 3. Experimental setup

Figure 1 shows the experimental setup used to determine the change in diameter of a nylon string. The main elements involved, that is, the He-Ne laser, the nylon string, the observation screen, and the hanging weight ($0.2 \text{ kg} \leq m \leq 1.7 \text{ kg}$) are included. Two photos of the actual experimental setup are shown in the lower panel of the figure. The diameter of the nylon string without any tensile applied was measured with a micrometre. The length from the string to the screen was measured with a measuring tape graduated in millimetres. The hanging weights consisted of calibrates masses where the uncertainty is provided by the manufacturer. We took high resolution pictures of the resulting diffraction pattern on the screen, at a fixed distance, so as to measure the positions of the diffraction minima with greater precision. The camera was placed at the front of the screen, on the same side as the laser, right in the middle below the laser beam, just facing perpendicularly the centre of the diffraction patterns. The pictures of the different patterns were taken with the camera at the same position. We located the positions of the minima by means of measuring the distance in pixels from the centre of the pattern. Then, a real measure was used to get back the distances in millimetres (Table 1).

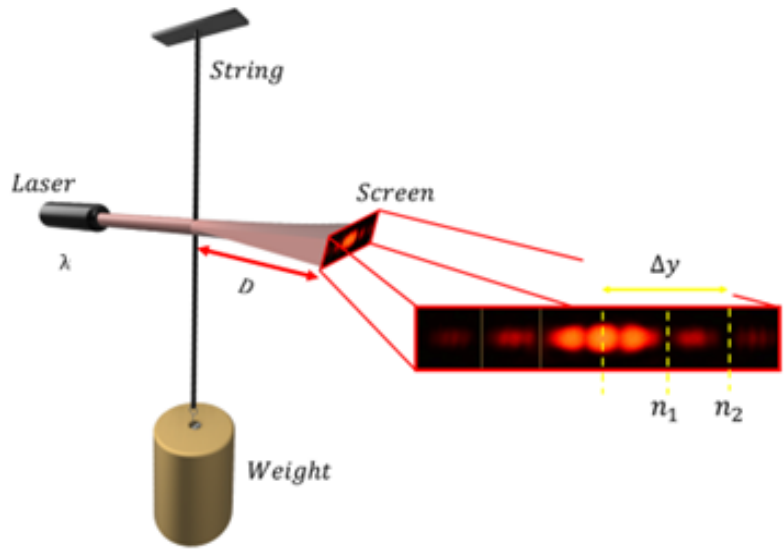

**Figure 1.** Schematic representation of the experimental setup used to determine the change in diameter of a nylon string when stretched using hanging weights. In the photograph, $n_1$ and $n_2$ indicate the diffraction minima positions and $\Delta y$ the distance to the center of the central peak.

## 4. Results and discussion

Figure 2 shows the photos of the diffraction spectra for the smallest (0.2 kg) and largest (1.7 kg) hanging weights. Two narrow minima can be observed inside the central peak of the spectra. This is a direct consequence of the Gaussian amplitude distribution of the wave field of the laser beam as proven in Ref. [15]. One can clearly observe the central diffraction pattern, as well as the first and second order diffraction patterns. Each order is separated from the next by a point where the diffracted intensity is minimum, and which corresponds to the region of destructive interference. The values for the hanging masses, the diffraction minima positions and the diameters have been included in Table 1. The procedure used to determine the position of the minima with respect to the center of the pattern using high resolution pictures was key as the corresponding uncertainty drives the uncertainty of the diameter as calculated using Eq. 6.

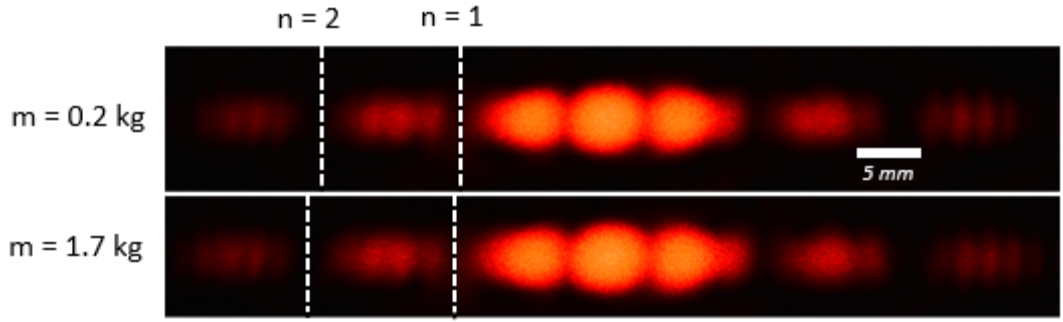

**Figure 2.** Diffraction spectra for the nylon string with $m = 0.2$ kg (top) and $m = 1.7$ kg (bottom). The position of the intensity minima for $n = 1$ and $n = 2$ are highlighted by the dashed lines. A scale bar representing 5 mm on the observation screen has been included in the picture.

The change in the distances of the diffraction minima from the center of the spectrum is well-resolved and can be used to calculate the diameter of the nylon string. For each value of the hanging mass, the diffraction minimum positions were measured and the corresponding diameters were calculated from Eq. (6). The result is shown in Figure 3.

Table 1. Mass $(m)$ of the hanging weights (calibrated masses are used), position $(\Delta y)$ of the $n = 2$ intensity minimum on the diffraction spectrum, and diameter $(d)$ calculated from Eq. 6. The uncertainties are included between parentheses.

| $m(kg)$ | Second minimum (mm) | $d = 2\lambda \frac{D}{\Delta Y}$ (mm) |
| --- | --- | --- |
| 0.20000 (0.00020) | 22.0905 (0.0058) | 0.257893 (0.000071) |
| 0.50000 (0.00035) | 22.3159 (0.0058) | 0.255471 (0.000069) |
| 0.80000 (0.00043) | 22.4390 (0.0058) | 0.253886 (0.000069) |
| 1.10000 (0.00052) | 22.5784 (0.0058) | 0.252321 (0.000068) |
| 1.40000 (0.00057) | 22.7177 (0.0058) | 0.250773 (0.000067) |
| 1.70000 (0.00064) | 22.9281 (0.0058) | 0.248743 (0.000066) |

Eq. (5) can be represented as $d = A + Bm$, where $A = d_0$ is the intercept with the ordinate and $B = \frac{4\nu g}{\pi d_0 E}$, the slope. Figure 3 also shows a linear fitting of $d$ versus $m$ from the data registered in Table 1. The fitting yields an intercept of $A = 2.589 \, (0.003) \times 10^{-4}$ m (which equals $d_0$) and a slope (absolute value) of $B = -6.1 \, (0.2) \times 10^{-6}$ m/kg. The coefficient of determination is $R^2 = 0.997$. The value of $d_0$ provided by the manufacturer for the diameter of the string is 0.26 mm, so we have obtained an excellent agreement. Using a Poisson's ratio of $\nu = 0.39 \, (0.02)$ [16], and $g = 9.80151 \text{ m/s}^2$ for the gravitational constant (value reported for Valencia city (Spain) [17]), we find a Young's modulus for the nylon string of $E = 3.1\ (0.2)\ \text{GPa}$. This value is also in excellent agreement with reported values for nylon: 1.59-3.79 GPa [16].

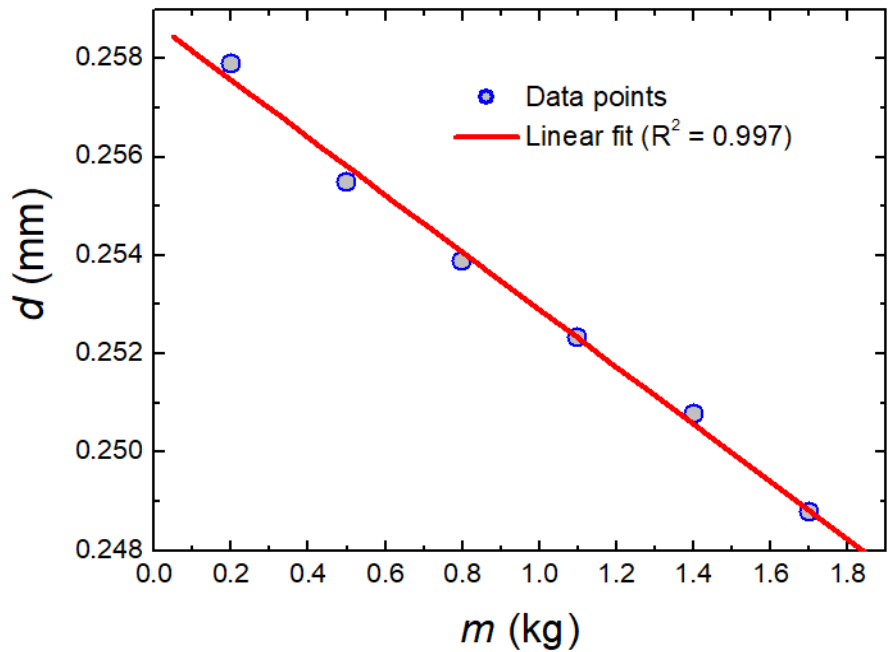

**Figure 3.** Diameter of the string as calculated from Eq. 6 as a function of the corresponding mass of the hanging weight.

A similar experiment with no need for including Poisson's ratio would require the measurement of the increments of length for each hanging weight. This experiment would take longer, this is why we would rather recommend it for a student project.

## Conclusions

This work shows a low-cost experiment to determine the Young's modulus of a nylon string. For this purpose, concepts from two different classes, namely, mechanics and optics are combined in a simple model. The change in the diameter caused by a hanging mass (elasticity theory) is measured by means of a diffraction experiment (wave optics). It therefore shows students that physics fields are often interconnected. The Young's modulus of the nylon string as determined from the model is in good agreement with the reported one. Our article provides a simple method involving laser measurements which can be adequate for first year university experiments.

## Acknowledgments

This work was supported by the Spanish Ministerio de Ciencia e Innovación (grant PID2022-142407NB-I00) and by Generalitat Valenciana (grant CIPROM/2022/30), Spain. The authors would like to thank the Instituto de Ciencias de la Educación (Institute of Education Sciences) at the Universitat Politècnica de València (Technical University of Valencia), Spain, for its support to the teaching innovation group MSEL.

## Disclosures

The authors have no conflicts to disclose.